\pdfoutput=1
\RequirePackage{amsmath}
\documentclass{llncs}
\usepackage{cite}
\usepackage{amssymb}
\usepackage{mathtools}
\usepackage{rgalg}
\usepackage{hyperref}
\usepackage{xcolor}
\usepackage{subfigure}
\usepackage{gnuplot-lua-tikz}

\def\miko{Mikol\'a\v{s} Janota}

\newcommand*\df[1]{\emph{#1}}  

\renewcommand*\vec{\mathaccent"017E}

\newcommand*\asgn{\mathrel{\coloneqq}}
\newcommand*\bmat[1]{\hat{#1}}
\newcommand*\defeq{\mathrel{\coloneqq}}
\newcommand*\liff{\leftrightarrow}
\newcommand*\limp{\to}

\newcommand*\cG{\mathcal{G}}
\newcommand*\ZZ{\mathbb{Z}}

\DeclareMathOperator*{\atleast}{\textsf{atleast}}
\DeclareMathOperator*{\atmost}{\textsf{atmost}}

\DeclareMathOperator{\less}{\textsf{less}}
\DeclareMathOperator{\lessbin}{\textsf{lex}}
\DeclareMathOperator{\lessunr}{\textsf{lessunr}}
\DeclareMathOperator{\unary}{\textsf{unary}}
\DeclarePairedDelimiter{\ceil}{\lceil}{\rceil}

\definecolor{darkblue}{rgb}{0,0,0.2}
\hypersetup{colorlinks,linkcolor=darkblue,citecolor=darkblue,urlcolor=darkblue}
\begin{document}
\title{On the Quest for an Acyclic Graph}
\author{\miko\inst{1}\! \and Radu Grigore\inst{2}\! \and Vasco Manquinho\inst{1}\!}
%
\institute{IST/INESC-ID, Lisbon, Portugal\\
\and School of Computing, University of Kent, UK}

\maketitle              

\begin{abstract}
    The paper aims at finding acyclic graphs under a given set of constraints.
 More specifically, given a propositional formula~$\phi$ over edges of a
 fixed-size graph, the objective is to find a model of~$\phi$ that corresponds to a
 graph that is acyclic. The paper proposes several encodings of the problem and
 compares them in an experimental evaluation using state-of-the-art SAT solvers.

\end{abstract}

\section{Introduction}

SAT solvers have become popular means of solving computationally hard problems.
However, most modern SAT solvers require  the  problem to be specified in
conjunctive normal form (CNF).  This may pose difficulty in problems comprising
of  constraints that do not have a straightforward translation to propositional
logic.  This paper targets one such constraint, \emph{graph acyclicity}.

A graph is naturally modeled by a set of  Boolean variables $x_{jk}$ expressing
the existence of an edge from $j$ to $k$.  So, for instance, a constraint
$x_{12}\lor x_{13}$ expresses that we're looking for  a graph where the
node 1 is connected to at least one of the nodes 2~and~3.  But how do we ensure that we are
looking for a graph that is acyclic?

Acyclicity is a core  concept from graph theory  and naturally arises in a
number of applications.  In planning it is used to ensure
causality~\cite{rintanen-ai06}.  Similarly, it is needed in networks
used for computation, e.g.\ Bayesian networks~\cite{vanbeek-cp15}.  More recently, software verification of concurrent programs relies on relaxed memory models, which are defined in terms of acyclic relations~\cite{lustig16,batty-popl17}.

There is a large body of work of translating constraints into CNF.  Prime
examples are cardinality constraints~\cite{Bailleux03,sinz-cp05,silva-cp07}
pseudo-Boolean constraints~\cite{een06,bailleux06},
XOR-constraints~\cite{kullmann-lata14} and general CSP~\cite{Tamura09}. A number
of graph-related problems approached by SAT can be found in the literature.  For
instance, encoding \emph{graph connectivness} property~\cite{tavares12}, graph
coloring~\cite{VanGelder-dam08}, calculating \emph{Steiner
tree}~\cite{Selman95,deOliveira14} or \emph{maximum
clique}~\cite{Ignatiev-ijcai07}.  It is also worth mentioning that graphs and
SAT play an important role in the research on \emph{Gene regulatory
networks}~\cite{dubrova11}.

The paper has the following contributions. It reviews CNF encodings that are
found in the literature, and introduces a number of new ones.  All these
encodings  were implemented and extensively evaluated.  To generate a
well-defined testbed we introduce The Supervisor Problem, which asks to assign
supervisors to a group of persons, where there is a lower bound for each person's
supervisors and an upper bound on how many persons he or she can supervise.
This problem is a generalization of the well-known ordering principle (also
known as the GT family) studied in proof complexity
\cite{krishnamurthy85,staalmarck96,bonet01,Alekhnovich-tc07}.


\section{Preliminaries}

Our basic setting is the following.
We represent a directed graph on $n\ge1$ vertices by Boolean variables~$x_{ij}$, associated with edges.
A family of graphs is defined by a formula $\phi_n(\vec{x},\vec{y}\,)$ over the edge variables~$\vec{x}$ and some auxiliary variables~$\vec{y}$.
Each graph corresponds to a satisfying assignment of $\exists\vec{y}\,\phi_n(\vec{x},\vec{y})$.
Given such a formula~$\phi_n$, we wish to construct another formula $\psi_n$, such that the satisfying assignments of $\exists\vec{y}\,\exists\vec{z}\,\bigl(\phi_n(\vec{x},\vec{y}\,)\land\psi_n(\vec{x},\vec{z}\,)\bigr)$ correspond to the graphs from the given family that, in addition, are acyclic.
We refer to the formula~$\psi_n$ as the \df{acylicity checker} or, for short, the \df{checker}.
In the rest of the section, we make this more precise, and we describe the particular formulas~$\phi_n$ that we use in experiments.
The acyclicity checkers $\psi_n$ are discussed later~(\autoref{sec:encs}).
Standard terminology from propositional logic and graph theory is assumed.

\subsection{Propositional Logic and SAT Solving}
\df{Propositional formulas} are constructed from propositional variables by Boolean
connectives ($\lor$, $\land$, $\Rightarrow$) and negation ($\lnot$).
A propositional formula is in \df{conjunctive normal form (CNF)} if
it is a conjunction of clauses, where a \df{clause} is a disjunction of
literals, where a \df{literal} is a  propositional variable or its negation.

Whenever convenient, we treat CNF as a set of clauses and a clause as a set of
literals.  Observe that the empty set of clauses corresponds semantically to
true and the empty clause corresponds semantically to false.
The empty clause is denoted as $\bot$.

Any propositional formula can be converted to an equisatisfiable CNF formula in linear time~\cite{tseitin68,plaisted-jsm86}.

The problem of satisfiability (SAT) is to decide for a given propositional formula whether
there is a satisfying assignments for it or not.

The paper requires minimal understanding of SAT solving.  We will assume that a
SAT solver accepts a formula in CNF and gives the response ``YES'' if the formula
is satisfiable and ``NO'' otherwise. Most modern SAT solvers also provide an
actual satisfying assignment if the answer is ``YES''.

When discussing the performance of SAT solvers,  sometimes we refer to the underlying proof system,
which is propositional resolution. Given two clauses $x\lor C$ and $\lnot x\lor D$
their resolution is $C\lor D$. The two clauses are called antecedents and the resulting clause the resolvent.
Resolution proof of a clause $C$ from a formula  $\phi$ is a sequence of
clauses $C_1,\dots,C_k$ such that $C_k=C$ and each clause $C_i$ is either in $\phi$
or it is a resolvent of some $C_j,C_k$ with $1\le j< k<i$.
A resolution refutation is a resolution proof of the empty clause.
A CNF formula is unsatisfiable if and only if there exists a resolution refutation for it.
For most modern SAT solvers it holds that for any run of the SAT solver that
responds ``NO'', there exists a resolution refutation of the input formula linear
to the run of the solver.  In fact, it even holds that for any resolution refutation
there exists a polynomial run of the solver~\cite{Darwiche-ai11}.

\subsection{Graphs}
Throughout the paper we will be discussing directed graphs.
A \df{directed graph} $G=(V,E)$ consists of a set~$V$ of \df{vertices} and a set~$E$ of \df{edges}.
In this paper, we work with finite graphs, and we fix $V=[n]=\{1,\ldots,n\}$.
We denote vertices by $i,j,k,\dots$ and we abbreviate the (directed) edge $(i,j)$ by~$ij$.
For an edge~$ij$, we say that it has \df{source}~$i$ and \df{target}~$j$;
  conversely, we say that $ij$~is an \df{outgoing} edge of~$i$, and an \df{incoming} edge of~$j$.
The \df{in-degree} of a vertex is its number of incoming edges,
  and the \df{out-degree} of a vertex is its number of outgoing edges.
A \df{sink} is a vertex with out-degree~$0$.

A \df{graph property} is a family~$\cG$ of graphs.
In this paper, we consider only properties~$\cG_n$ that contain graphs with a fixed number~$n$ of vertices, and are defined by a formula~$\phi_n$.
More precisely, $\phi_n$~has one variable $x_{ij}$ for each pair $(i,j)\in V\times V$, and may also use some auxiliary variables~$\vec{y}$.
Then, a graph with edges~$E$ is in $\cG_n$ if and only if it corresponds to a satisfying assignment of $\exists\vec{y}\,\phi_n(\vec{x},\vec{y})$.
We say that a property is \df{monotone} if adding edges to a graph preserves the property.
This corresponds to the Boolean function $\exists\vec{y}\,\phi_n(\vec{x},\vec{y}\,)$ being monotone. 
We remark that the property of having a cycle is monotone, and the intersection of two monotone properties is monotone.

\subsection{The No-Sink Family}

The graphs with no sinks are described by the following formula:
\begin{align}
  \phi_n(\vec{x}\,)\;\defeq\;
    \bigwedge_{i \in [n]} \biggl( \bigvee_{j\in [n]} x_{ij}\biggr)
  \tag{\textsf{no-sink}}
  \label{eq:no-sink}
\end{align}
Observe that the no-sink property is monotone.
Also, the \df{ordering principle} says that graphs with no sink have a cycle.
It follows that, no matter what acyclicity check~$\psi_n$ we use, the formula $\exists\vec{y}\,\exists\vec{z}\,\bigl(\phi_n(\vec{x},\vec{y}\,)\land\psi_n(\vec{x},\vec{z}\,)\bigr)$ will be unsatisfiable.
Since we know the answer, the no-sink family is a good test case for the correctness of our implementation.
In addition, as we shall see, these unsatisfiable formulas are not easy for state-of-the art SAT solvers.

%
%
%

\subsection{The Supervisor Problem}

We define a parameterized family of graphs, which will typically contain both cyclic and acyclic graphs.
The idea is as follows. Consider a group of $n$~people
where each person must have a certain number of supervisors and at the same
time there is a limit on how many people a person can supervise.
The supervisor problem is formalized as
follows.

\medskip
\noindent \textbf{Input:} A positive  integer $n$ and a sequence of pairs of nonnegative integers $(u_1,l_1),\dots, (u_n,l_n)$.
\par\nobreak
\noindent \textbf{Output:} ``YES'' if and only if there exists an acyclic graph  with the vertices $V=[n]$ such that vertex~$i$ has at most $u_i$ incoming edges and at least $l_i$ outgoing edges.
\medskip


An instance $n,\vec{u},\vec{l}$ of the Supervisor problem reduces to the satisfiability of the following formula:
\begin{align}
  \begin{aligned}
  \phi_{n,\vec{u},\vec{l}\,}(\vec{x}\,) \;\defeq\;
  &\bigwedge_{\mathclap{i\in[n]}} \atleast(l_i; x_{i1},\ldots,x_{in}) \\
  &{}\land \bigwedge_{\mathclap{j\in[n]}} \atmost(u_j; x_{1j},\ldots,x_{nj})
  \end{aligned}
  \tag{\textsf{supervisor}}
  \label{eq:supervisor}
\end{align}
The cardinality constraint $\atleast(l;\vec{x}\,)$ evaluates to~$1$ when at least~$l$ of the Boolean variables~$\vec{x}$ evaluate to~$1$.
There are several ways to encode such cardinality constraints in CNF~\cite{Bailleux03,sinz-cp05,een06,silva-cp07}.
The cardinality constraint $\atmost(u;\vec{x}\,)$ is analogous.

The Supervisor problem has three interesting special cases:
  (a)~the no-sink family;
  (b)~the pigeonhole principle; and
  (c)~the DAG realizability problem.
We obtain a description of the no-sink family by requiring the bounds $(n,1)$ for each of the $n$~vertices.
We obtain a description of the pigeonhole principle by requiring the bounds $(0,1)$ for $k_1$~vertices and the bounds $(1,0)$ for $k_2$~vertices, with $k_1+k_2=n$ and $k_1>k_2$.
Finally, any instance of the DAG realizability problem is also an instance of the Supervisor problem, which satisfies the additional constraint $\sum_{i=1}^n u_i = \sum_{i=1}^n l_i$.

Since the pigeonhole principle is a special case, it follows that \eqref{eq:supervisor} sometimes requires large resolution refutations~\cite{haken85}.
Since the DAG realizability problem is a special case, it follows that it is NP-complete to find a model of~\eqref{eq:supervisor} that corresponds to an acyclic graph~\cite{hartung-siam15}.
Finally, we observe that \eqref{eq:supervisor} is not monotone.

\section{Encodings}\label{sec:encs}

We discuss several ways of constructing the acyclicity checker~$\psi_n$,
  based on reachability~(\autoref{sec:tc}),
  based on labeling vertices with numbers (\autoref{sec:binary}~and~\autoref{sec:unary}),
  and based on oblivious algorithms~(\autoref{sec:fw}~and~\autoref{sec:mm}).

\subsection{Transitive Closure}\label{sec:tc}

A graph $G=(V,E)$ does not have cycles if and only if $E^+$ is irreflexive.
Here, $E^+$ denotes the irreflexive transitive closure of~$E$.
Further, $G$~does not have cycles if and only if there exists an irreflexive transitive relation~$R$ that contains~$E$; that is, if there exists $R\subseteq V\times V$ such that
(a)~$ii\notin R$ for all $i\in V$,
(b)~$ij\in R$ and $jk\in R$ only if $ik\in R$ for all $i,j,k\in V$, and
(c)~$ij\in R$ if $ij\in E$ for all $i,j\in V$.
We represent the relation~$R$ by variables $y_{ij}$, and we define $\psi_n$ by clauses that directly correspond to the conditions (a), (b), and~(c):
\begin{align}
  \psi_n(\vec{x},\vec{y}\,)\;\defeq\; 
    \bigwedge_{i\in[n]} \lnot y_{ii} 
    \land \bigwedge_{i,j,k\in[n]}
      (y_{ij} \land y_{jk} \limp y_{ik}) 
    \land \bigwedge_{i,j\in[n]}
      (x_{ij} \limp y_{ij})
  \tag{\textsf{tc1}}
  \label{eq:tc-trans}
\end{align}
A possible alternative is the following:
\begin{align}
  \psi_n(\vec{x},\vec{y}\,)\;\defeq\;
    \bigwedge_{i\in[n]} \lnot y_{ii}
    \land \bigwedge_{i,j,k\in[n]}
      (y_{ij} \land {\color{red}x_{jk}} \limp y_{ik})
    \land \bigwedge_{i,j\in[n]}
      (x_{ij} \limp y_{ij})
  \tag{\textsf{tc2}}
  \label{eq:tc-cyc}
\end{align}
In relational language, \eqref{eq:tc-trans} corresponds to the conditions (a)~$R\cap\mathsf{id}=\emptyset$, (b)~$RR\subseteq R$, and (c)~$E\subseteq R$.
Here, $\mathsf{id}$ is defined to be $\{\,ii\mid i\in V\,\}$.
In~\eqref{eq:tc-cyc}, we replace condition~(b) by $RE\subseteq R$.
Juxtaposing relations corresponds to normal relational composition:
\begin{align*}
RE \;\defeq\; \{\,(i,j)\mid\text{$(i,k)\in R$ and $(k,j)\in E$, for some $k$}\,\}
\end{align*}

If the family of graphs satisfying~$\phi_n$ is monotone, in the sense that adding edges preserves~$\phi_n$, then we do not need to use auxiliary variables $y_{ij}$, as we can simply reuse the variables~$x_{ij}$:
\begin{align}
  \psi_n(\vec{x}\,)\;\defeq\;
    &
    \bigwedge_{i\in[n]} \lnot x_{ii}
    \land \bigwedge_{i,j,k\in[n]} (x_{ij} \land x_{jk} \limp x_{ik})
  \tag{\textsf{tc3}}
  \label{eq:tc-mono}
\end{align}
The conjunction of \eqref{eq:tc-mono}~and~\eqref{eq:no-sink} has $\Omega(2^n)$ regular tree resolution proofs but $O(n^3)$ dag resolution proofs; see~\cite{bonet01,krishnamurthy85,staalmarck96} and \cite[\S7.2.2.2, Theorem~R]{KnuthACP6}.
Such conjunctions express the \df{ordering principle}.

Observe that the formulas in each of \eqref{eq:tc-trans}, \eqref{eq:tc-cyc}, and~\eqref{eq:tc-mono} have size~$\Theta(n^3)$.

\subsection{Binary Labeling}\label{sec:binary}

A graph is acyclic when it can be embedded in a linear order;
  that is, when there exists a labeling $l:V\to\ZZ$ of the vertices such that $ij\in E$ implies $l(i)<l(j)$ for all $i,j\in V$.
Recall that $V=[n]$.
Clearly, it is sufficient to use $b \defeq \lceil \log_2 n\rceil$ bits for labels, so $l:V\to\{0,1\}^b$.
Accordingly, we use $n$~groups $\vec{y}_1,\ldots,\vec{y}_n$ of auxiliary variables, each containing $b$~Boolean variables.
The acyclicity checker is then~\cite{rintanen-ai06}:
\begin{align}
  \psi_n(\vec{x},\vec{y}_1,\ldots,\vec{y}_n) \;&\defeq\;
    \bigwedge_{\mathclap{i,j\in[n]}}
      \bigl(x_{ij} \limp \less(\vec{y}_i,\vec{y}_j) \bigr)
  \tag{\textsf{bin}}
  \label{eq:bin}
\end{align}
The implementation of $\less$ depends on how we represent numbers.
Because we use binary numbers, we set $\less \defeq \lessbin_b$, and define the latter to test lexicographic ordering of the bits:
\begin{align}
  \lessbin_0() \;&\defeq\; 0
\\
  \lessbin_b(\vec{y}y,\vec{z}z) \;&\defeq\;
    (\lnot y \land z) \lor
      \bigl( (\lnot y \lor z) \land \lessbin_{b-1}(\vec{y},\vec{z}\,) \bigr)
\end{align}
Converting this circuit to CNF will require some more auxiliary variables, which we do not denote explicitly.
The conversion is standard, and the size of the resulting CNF is $\Theta(b)$.

Observe that the formula in~\eqref{eq:bin} has size $\Theta(n^2 \log n)$.

\subsection{Unary Labeling}\label{sec:unary}

We can also use the labeling idea with unary numbers.
To represent a number, instead of $\ceil{\log_2 n}$ bits we use $n-1$~bits, and force them to follow the pattern $0^*1^*$.
We could still compare these numbers by their lexicographic order using~$\lessbin$, but we can also write directly a CNF formula, thus avoiding the (small) overhead associated with converting a circuit into CNF\null.

The acyclicity checker is defined by
\begin{align}
  \psi_n(\vec{x},\vec{y}_1,\ldots,\vec{y}_n\,) \;&\defeq\;
    \bigwedge_{\mathclap{i,j\in[n]}}
      \bigl(x_{ij} \limp \less(\vec{y}_i,\vec{y}_j) \bigr)
    \land
    \bigwedge_{\mathclap{i\in[n]}}
      \unary(\vec{y}_i)
  \tag{\textsf{unr}}
  \label{eq:unr}
\end{align}
where each $\vec{y}_i$ uses $n-1$~bits.

We could use $\less \defeq \lessbin_{n-1}$, but we prefer $\less(\vec{y},\vec{z}\,) \defeq \exists\vec{u}\,\lessunr(\vec{y},\vec{z},\vec{u}\,)$, where
\begin{align}
  \lessunr(\vec{y},\vec{z},\vec{u}\,) \;&\defeq\;
    \bigwedge_{i=1}^{n-1}
      \bigl(
        (\lnot y_i\lor \lnot u_i)
        \land
        (z_i \lor \lnot u_i)
      \bigr)
    \land
    \bigvee_{i=1}^{n-1} u_i
    \label{eq:unrless}
\end{align}
The auxiliary $\vec{u}$ variables will identify a position~$i$ where $y_i=0$ and $z_i=1$.
This is sufficient because we enforce the pattern~$0^*1^*$.
\begin{align}
  \unary(\vec{y}\,) \;&\defeq\;
    \bigwedge_{i=2}^{n-1} (y_{i-1} \limp y_i)
\end{align}

Observe that the formula in~\eqref{eq:unr} has size $\Theta(n^3)$.

\subsection{Warshall}\label{sec:fw}

The Warshall algorithm computes transitive closure, in cubic time.
\medskip
{\sffamily
\begin{alg}
\^ \proc{Warshall}
\0  ~for~ $k\in[n]$
\1    ~for~ $i\in[n]$
\2      ~for~ $j\in[n]$
\3        $a_{ij} \asgn \mathsf{Or}(a_{ij}, \mathsf{And}(a_{ik},a_{kj}))$
\end{alg}}
\medskip\noindent
Initially, $a_{ij}$ says if the graph contains an edge~$ij$;
  and, at the end $a_{ij}$ says if the graphs contains a path from~$i$ to~$j$.
In particular, at the end, $a_{ii}$ says if there is a cycle containing vertex~$i$.

In a usual implementation, the procedures $\mathsf{Or}$, $\mathsf{And}$ are simply performing the corresponding Boolean operations.
Instead, we could replace each of them with a function that constructs the corresponding circuit.
Because the Warshall algorithm is simple enough, we can also build the acyclicity checker directly.
We use $n^2(n+1)$ auxiliary variables $y_{ijk}$, for $i,j\in[n]$ and $k\in\{0,\ldots,n\}$.
The variable $y_{ijk}$ will be~$1$ only if there exists a path from~$i$ to~$j$ that uses intermediate vertices only from the set~$[k]$.
\begin{align}
\begin{aligned}
  \psi_n(\vec{x},\vec{y}) \;\defeq\;
  &
  \bigwedge_{i\in[n]} \lnot y_{iin} 
  \land \bigwedge_{i,j\in[n]} (x_{ij}\limp y_{ij0})
  \land \bigwedge_{i,j,k\in[n]} (y_{ij(k-1)}\limp y_{ijk}) \\
    &{}\land \bigwedge_{i,j,k\in[n]}
      (y_{ik(k-1)}\land y_{kj(k-1)}\limp y_{ijk}) \\
\end{aligned}
  \tag{\textsf{fw}}
  \label{eq:fw}
\end{align}
We remark that the formula above is close but \emph{not} the same as what we would obtain by replacing the basic operations $\mathsf{Or}$,~$\mathsf{And}$ with circuit builders.
One difference is that we use implications rather than equivalences: It is possible that $y_{ijk}=1$ even if there is no corresponding path.
Another difference is that the last conjunct built by Warshall would be
$
  (y_{ik(k-[k\ge j])}\land y_{kj(k-[k\ge i])}) \liff y_{ijk}
$,
where $[p]$ evaluates to~$1$ when $p$~is true, and $0$ when $p$~is false.
Either form is correct.

The formula in~\eqref{eq:fw} is similar to the one in~\eqref{eq:tc-trans}.
It has the same size $\Theta(n^3)$, but uses more auxiliary variables.

\subsection{Matrix Multiplication}\label{sec:mm}

Let us take stock of the encodings we have so far.
Four encodings have size $\Theta(n^3)$, namely \eqref{eq:tc-trans}, \eqref{eq:tc-cyc}, \eqref{eq:unr}, and~\eqref{eq:fw}; another encoding has size $\Theta(n^2 \log n)$, namely~\eqref{eq:bin}.
But, size is not the only criterion on which to judge a satisfiability formula.
We note that \eqref{eq:tc-trans}, \eqref{eq:tc-cyc}, and~\eqref{eq:fw} have the following desirable property: \emph{if the edge variables~$\vec{x}$ are fixed, then satisfiability is decided by unit propagation}.
The smaller encoding \eqref{eq:bin} does not have this property.
Is it possible to have a formula of sub-cubic size that is solved by unit propagation once the edge variables~$\vec{x}$ are fixed?
We answer this question affirmatively by showing an encoding based on matrix multiplication.
We first describe the algorithm.
The formula~$\psi_n$ will be constructed by replacing the basic operations with circuit builders, and then converting the circuit to CNF\null.

We will consider $n\times n$ matrices of two kinds, over integers and over Booleans.
Integers form a commutative ring, so the corresponding matrix operations are unrestricted.
Booleans do not form a ring:
We take addition to be logical-or and we take multiplication to be logical-and, but there is no subtraction.
Since we use both types of operations and sometimes reinterpret the same matrix as being over integers or over Booleans, we distinguish Boolean matrices with a hat.
When we reinterpret an integer matrix~$X$ as a Boolean matrix~$\bmat{X}$, all non-zero integers collapse to~$1$.

Let $X$~be the adjacency matrix of a graph, with entries in $\{0,1\}$.
Entry~$(i,j)$ of~$X^l$ is the number of paths of length~$l$ from~$i$ to~$j$, and entry $(i,j)$ of~$\bmat{X}^l$ tells us if there exists a path of length~$l$ from~$i$ to~$j$.
Thus, to decide if the graph has a cycle it suffices to check the diagonal of the Boolean matrix $\sum_{l=1}^n\bmat{X}^l$.
We compute this matrix by repeated squaring:
\begin{align}
  \bmat{A}_0 &\defeq \bmat{X}
&
  \bmat{A}_{k+1} &\defeq \bmat{A}_k^2
\\
  \bmat{B}_0 &\defeq \bmat{X}
&
  \bmat{B}_{k+1} &\defeq \bmat{B}_k + \bmat{A}\bmat{B}_k
\end{align}
It is easy to check that $\bmat{A}_k=\bmat{X}^{2^k}$ and $\bmat{B}_k=\sum_{l=1}^{2^k}\bmat{X}^l$, and also that $\bmat{B}_k$ equals $\sum_{l=1}^n \bmat{X}^l$ whenever $k\ge\log_2 n$.
Thus we can decide if a directed graph is acyclic in the time needed to perform $\Theta(\log n)$ multiplications of $n\times n$ Boolean matrices.
One could perform a naive matrix multiplication in $\Theta(n^3)$ time, giving an acyclicity checker $\psi_n$ of size $\Theta(n^3 \log n)$.

Alternatively, to multiply Boolean matrices $\bmat{A}$~and~$\bmat{B}$, we multiply them as integer matrices obtaining $C=AB$; the product of $\bmat{A}$~and~$\bmat{B}$ is then~$\bmat{C}$.
(The idea of using intermediate integer matrices is old~\cite{fisher71,furman70,munro71}.)
To perform the multiplication of integer matrices, we use Strassen's algorithm~\cite{strassen-matmul}, which works in $O(n^{2.81})$ time.
To achieve this improved runtime, Strassen's algorithm makes essential use of subtraction, which is not available over Booleans.
However, the runtime $O(n^{2.81})$ counts operations over integers, while we must count operations over bits since we ultimately want to build a circuit.
Since the matrices $A$~and~$B$, which we multiply, have entries from $\{0,1\}$, the product matrix~$C$ has entries from~$\{0,\ldots,n\}$.
Some intermediate values in Strassen's algorithm are outside this range.
Nevertheless, one can check that $\Theta(\log n)$ bits are sufficient to represent all intermediate values.
So, the size of the circuit is $O(n^{2.81} \log n \cdot f(\log n))$, where $f(w)$ is the size of a circuit necessary to perform an arithmetic operation on integers with $w$~bits.
In our implementation, we use a simple quadratic multiplication algorithm, leading to a circuit of size $O(n^{2.81}\log^3 n)$.

In theory, one can use the same approach to achieve $O\bigl(n^{\omega}\log^2n\log\log n\cdot 8^{\log^*\log n}\bigr)$ by using better algorithms for the multiplication of matrices~\cite{legall:issac14} and for the multiplication of integers~\cite{harvey:jc2016}.
But such algorithms are known to not be practical.
Our current implementation based on Strassen is not competitive either~(\autoref{sec:eval}). 
Improving it seems to require new ideas.


\paragraph{Discussion.}

The approach we took for both Warshall and for matrix multiplication was to replace the basic operations in the algorithm with circuit builders.
This approach is fairly general, but it does not apply to all algorithms, just to the oblivious ones.
An algorithm is said to be \df{oblivious} when the memory locations it accesses do not depend on the input data.
If an algorithm runs in $t(n)$ time in the Turing model of computation, then one can construct an oblivious algorithm that runs in $O\bigl(t(n)\log t(n)\bigr)$ time, and hence a Boolean circuit of that size~\cite{pippenger:jacm79}.
In contrast, a similar construction does \emph{not} exist in the deterministic RAM model of computation~\cite{goldreich:jacm96}.

The standard (and optimal) algorithm for finding cycles in directed graphs is DFS (depth-first search).
However, we are not aware of an implementation of DFS that works in sub-cubic time in the (oblivious, deterministic, multi-tape) Turing machine model of computation.

\section{Experimental Evaluation}\label{sec:eval}

\begin{figure}\centering
\input{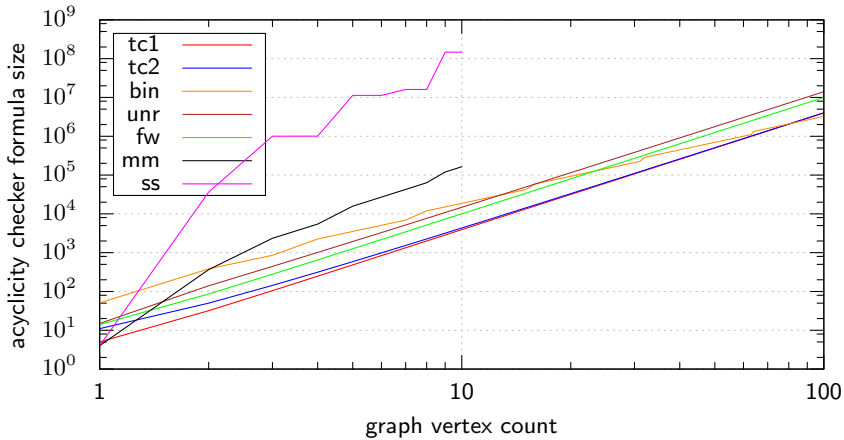}
\caption{
  The size of the acylicity checker formula~$\psi_n$ as a function of the number of vertices~$n$.
  Five lines correspond to equations \eqref{eq:tc-trans}, \eqref{eq:tc-cyc}, \eqref{eq:bin}, \eqref{eq:unr}, and~\eqref{eq:fw}.
  The remaining two correspond to Boolean matrix multiplication methods, using the definition (\textsf{mm}), and using Strassen's algorithm~(\textsf{ss}).
  The size of a CNF formula consisting of clauses $C_1,\ldots,C_m$ is defined to be $m+\sum_{i=1}^m |C_i|$.
}
\label{fig:formulasize}
\end{figure}

\paragraph{Overall setup.}
Three popular state-of-the-art SAT solvers were used for the evaluation:
Lingeling Version 276~\cite{biere13}, glucose 4.0~\cite{audemard09}, and minisat~2.2 (from github)~\cite{Een03}.
Further, minisat was a run in two modes: the default mode, and with
additional preprocessing techniques enabled (asymmetric breaking and redundancy checking).
The encodings we consider are Transitive closure~\eqref{eq:tc-trans}, Unary labeling~\eqref{eq:unr},
Binary labeling~\eqref{eq:bin}, Warshall~\eqref{eq:fw}, and an alternate version of Transitive closure~\eqref{eq:tc-cyc}.
The techniques based on matrix multiplication are not included, because they quickly lead to formulas in the realm of hundred of megabytes, and more (\autoref{fig:formulasize}).

The experimental results were obtained on an Intel Xeon 5160 3GHz,
with 4GB of memory, time limit $500$\,seconds, and memory limit~2GB.

\paragraph{Generation of Graph Families.}

The generation of the no-sink graph family is trivial: we simply instantiate~\eqref{eq:no-sink} for different values of~$n$.
We also generate random instances of the Supervisor problem.
Each upper bound~$u_i$ on the in-degree of vertex~$i$ is chosen uniformly at random from the set $\{0,\ldots,n-1\}$.
Each lower bound~$l_i$ is~$0$ with probability~$p$, which is a parameter, and otherwise is chosen uniformly at random from the set $\{1,\ldots,n-1\}$.
If the generated sequence $(l_1,u_1),\ldots,(u_n,l_n)$ is not satisfied by any directed graph, we throw it away and try again.
This filtering is easy, because it is possible to check in linear time whether a sequence is satisfiable by some directed graph, using the Fulkerson--Chen--Anstee theorem~\cite{anstee:dm82}; but, as we mentioned, checking whether a sequence is satisfiable by an \emph{acyclic} directed graph is NP-complete~\cite{hartung-siam15}.

For each $n\in\{2,3,\ldots,50\}$ and for each $p\in\{10\%,20\%,\ldots,90\%\}$, we generate a sequence $(l_1,u_1),\ldots,(l_n,u_n)$ using the method just described.
For each such sequence, we generate the formula~\eqref{eq:supervisor}.
The $\atmost$ and $\atleast$  constraints were encoded into CNF using the well-known
\emph{Sequential Counter} encoding~\cite{sinz-cp05}.

\begin{figure}[t]
  \centering
\begin{minipage}{.4\textwidth}
  \subfigure[GT Family with 49 instances\label{fig:numbers:gt}]{
      \begin{tabular}{|c|c|c|c|c|c|}
          \hline
          \textbf{solver}/\textbf{checker}& \ref{eq:tc-trans} & unr & bin & fw & \ref{eq:tc-cyc} \\\hline\hline
          lingeling                       & \textbf{49} & 48 & 10 & 38 & 11\\\hline
          glucose                         & 32 & 37 & 14 & 15 & 11\\\hline
          minisat                         & 25 & \textbf{49} & 12 & 12 & 11\\\hline
          minisat-prepro                  & 26 & \textbf{49} & 11 & 19 & 11\\\hline
      \end{tabular}
  }
\end{minipage}
\hspace{10pt}
\begin{minipage}{.4\textwidth}
  \subfigure[Supervisor Family with  441 instances\label{fig:numbers:sp}]{
      \begin{tabular}{|c|c|c|c|c|c|}
          \hline
          \textbf{solver}/\textbf{checker}& \ref{eq:tc-trans} & unr & bin & fw & \ref{eq:tc-cyc} \\\hline\hline
          lingeling                       & 436 & 429 & 426 & 435 & 435\\\hline
          glucose                         & \textbf{437} & 434 & 427 & \textbf{437} & \textbf{437}\\\hline
          minisat                         & 435 & 425 & 424 & 435 & 435\\\hline
          minisat-prepro                  & 435 & 426 & 422 & 435 & 435\\\hline
      \end{tabular}
  }
\end{minipage}
    \caption{Number of solved instances for sizes 2-50.\label{fig:numbers}}
\end{figure}
\begin{figure}[t]
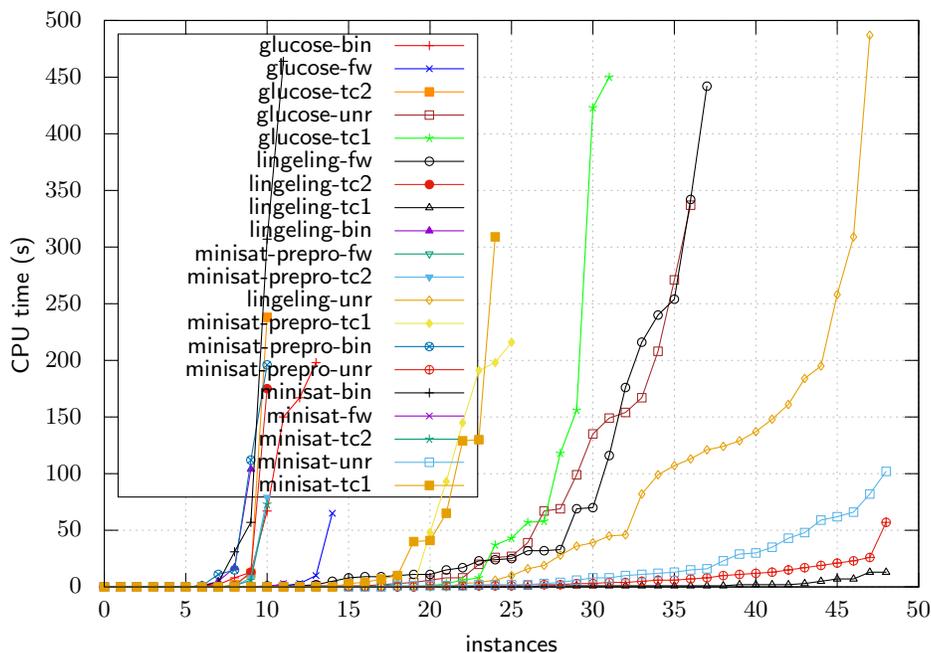

    \include{p-gt-all}
    \caption{Cactus plot for the \eqref{eq:no-sink} family.\label{fig:cactus}}
\end{figure}

\paragraph{Presentation of the results.}
Figure~\ref{fig:numbers} presents the number of instances solved.
Figure~\ref{fig:numbers:gt} shows the results for graphs with no sinks, which lead to the GT~family of formulas when coupled with the acyclicity checker~\eqref{eq:tc-trans}.
Figure~\ref{fig:numbers:sp} shows the results for  the general supervisor problem.
Figure~\ref{fig:cactus} shows a cactus for the \eqref{eq:no-sink} family for all the considered combinations of solvers and encodings.
Recall that if a cactus plot contains a point $(i, s)$
it means there are $i$~instances such that each was solved within $s$~seconds.
(Therefore, the graph is necessarily increasing.)
A more detailed presentation of the results can be found on the authors' homepage.\footnote{\url{http://sat.inesc-id.pt/~mikolas/sat-cycles/}}

The first thing to observe is that the supervisor family turns out to be easy for all the approaches.
Indeed, all approaches solved over $90\%$~benchmarks.
Even though the differences are rather small, the tables suggests that
the Transitive closure encoding~(\eqref{eq:tc-trans}) is good for all solvers.
In contrast, Binary labeling~\eqref{eq:bin} appears to be the worst.

In the case of \eqref{eq:no-sink} graphs, the results are widely different for each encoding but also for each solver.
Most notably, lingeling performs extremely well on~\eqref{eq:tc-trans}.
Both versions of minisat perform well on Unary labeling~\eqref{eq:unr}.
Glucose also performs the best with~\eqref{eq:unr}, but once the graph size gets over $30$~nodes, the performance quickly deteriorates.
The worst encoding appears to be Binary labeling~\eqref{eq:bin} and the alternative Transitive closure~\eqref{eq:tc-cyc}, with which all solvers could only handle less than half of all instances.
Warshall has a peculiar behavior, as it performs poorly for all the solvers except lingeling.

\paragraph{Discussion.}
 The results are surprisingly uneven across different solvers.
 In a particular, lingeling performs extremely well with~\eqref{eq:tc-trans},
 while the other solvers do not perform very well with the same encoding.
It is possible that lingeling benefits from some simplification technique (pre- or in-processing) that is missing from the other solvers, but this remains to be understood.
Binary labeling~\eqref{eq:bin} is overall the most compact but with the poorest performance.
This is not entirely surprising.
First, binary labeling does not offer many opportunities for applying unit propagation.
And second, binary labeling makes it virtually impossible to learn anything about any particular variable.

 In contrast, Unary labeling~\eqref{eq:unr} fares quite well over the considered solvers.
 Further, we observed that minisat with preprocessing solves the \eqref{eq:no-sink} family with no conflicts, i.e.,
 merely with preprocessing. This also holds for lingeling but only up to a certain size of the graph.
 It is not difficult to see how these formulas can be completely solved by preprocessing,
 in particular by  failed literal detection and asymmetric breaking.

 In unary labeling~(\autoref{sec:unary}), the ordering of labels is ensured by
the following constraint.  For every vertex~$i$ labeled by a number $l(i)\in\{0,\ldots,n-1\}$
there must be a neighbor~$j$  labeled by a number $l(j)\in\{0,\ldots,n-1\}$ such that $l(i)<k\le l(j)$
for some $k\in\{0,\ldots,n-1\}$.  To show that such a labeling does not exist for no-sink
graphs, we reason as follows.  If $k=n-1$,  it immediately follows that vertex~$j$ has  no neighbor labeled with a greater number since it has reached
the maximum.  Performing this reasoning for every vertex  yields that the
labels must be in the interval $\{0,\ldots,n-1\}$.
Repeating this reasoning $n$~times leads to a contradiction.

    We argue that to perform this reasoning, it is sufficient to use unit
    propagation on unary labeling.  In  definition~\eqref{eq:unrless}, setting
    $u_k\defeq 1$ imposes exactly the condition as listed above, i.e., the
    label of the source node is at most $k-1$ and the target node at least $k$.  It
    is easy to verify that setting $u_{n-1}\defeq 1$ yields a conflict by unit propagation.

    This idea is realized in the preprocessing techniques failed literal
    detection~\cite{crawford93,lynce-03} and asymmetric
    breaking~\cite{piette08-ecai08}. More generally, if setting a literal~$l$
    to true yields a conflict by unit propagation it is called a failed literal
    and the literal is set to false.
    In this manner, these two techniques acting in tandem can gradually remove the $u_i$~literal corresponding to the highest number,
    and eventually derive an empty clause.

\paragraph{Conclusion.}
   The experimental results support the following conclusions.
   The combination of choice for checking acyclicity is lingeling with Transitive closure~\eqref{eq:tc-trans}.
   (Do \emph{not} use the seemingly very similar alternative Transitive closure~\eqref{eq:tc-cyc}, which leads to poor performance.)
   Minisat is the second solver of choice but \emph{not} with the \eqref{eq:tc-trans} encoding---instead, use~\eqref{eq:unr}.
SAT solvers appear to be good at finding an acyclic graph if it exists; proving that an acyclic graph does not exist can easily become very difficult.

\paragraph{Formula Size.}

\autoref{fig:formulasize} presents the size of the acyclicity checkers~(\autoref{sec:encs}).
The cubic methods (\textsf{tc1}, \textsf{tc2}, \textsf{unr}, \textsf{fw}) are all within a factor of~$3$ of each-other.
The super-cubic one (\textsf{mm}, which is based on the definition of Boolean matrix multiplication) is already worse than the cubic ones at $n=3$, and it only gets worse for larger~$n$.
There are two sub-cubic methods: \textsf{ss} with size $O(n^{2.81}\log^3 n)$, and \textsf{bin} with size $O(n^2\log n)$.
The method based on Strassen (\textsf{ss}) generates large formulas even for $n=10$ and, looking at \autoref{fig:formulasize}, there is no overtaking in sight.
Indeed, even if the constants hidden in the $O$-notation were just as good for \textsf{ss} as they are for \textsf{tc1}, the crossing point would be at~$n\sim10^{32}$.
The $\log^3n$ factor is important, and it seems that without new ideas this method will never be practical.
On the other hand, the method based on binary labeling (\textsf{bin}), generates the smallest acyclicity checkers for $n\ge100$.

\section{Summary and Future Work}

The paper presents a number of encodings for the graph acyclicity constraint into conjunctive normal form.
This is useful when calculating acyclic graphs using SAT solvers but also other logic-based solvers, such as
quantified Boolean formula (QBF) and Satisfiability modulo theories (SMT) solvers.
The experimental evaluation suggests that the performance is highly sensitive both to be encoding but also to the SAT solver used.
This suggests that for practical purposes it is useful to maintain  a portfolio of encodings and solvers.

There are a number of directions for further research.
All the encodings that are asymptotically better than cubic perform poorly in practice.
Hence, it is a challenge for future research to identify encodings
with asymptotic complexity better than cubic but that also perform well.

Ultimately, all the considered encodings are not feasible for large graphs.
Indeed, any cubic encoding generates formulas that are  too large
once the number of nodes gets over 1000.  A possible solution is to consider
a lazy approach where only some portion of the encoding is considered.
However, this will probably require an iterative sequence of calls to the SAT
solver.
Another approach, tried recently~\cite{janhunen-fi16,cuteri-iclp17}, is to integrate
specialized algorithms for acyclicity inside the SAT solver.  This is also
closely related to \emph{loop formulas}, which are used in Answer Set
Programming (ASP) to avoid cyclic reasoning
cf.~\cite{lifschitz06,Janhunen11,gebser-jelia14}.

Some questions about properties of the encodings with respect to \emph{unit
propagation}  remain open.  In particular, is it the case that any partial
assignment that contains a cycle produces a conflict by unit propagation in a
given encoding?  In theory this is possible due to
monotonicity~\cite{Narodytska-09} (adding edges preserves cycles).  We
believe this holds for the encodings Transitive closure,
Warshall, and matrix multiplication, but does not hold for Unary and Binary labeling.
Hence, this also begs the question whether the labeling methods can be  made more deterministic.
Other properties related to unit propagation are also worth exploring, e.g.,
\emph{generalized arc consistency} and \emph{propagation completeness}, cf.~\cite{gent02,Narodytska-09,bordeaux12,cepek13,kullmann-lata14}.

The experimental results lead to several interesting questions.
Why is lingeling so  efficient on the transitive closure encoding ~\eqref{eq:tc-trans}?
Why  does the alternative encoding for transitive closure~\eqref{eq:tc-cyc}
perform so poorly across all solvers?
Do formulas~\eqref{eq:tc-cyc} lack short resolution refutations?

Last but not least, this research  opens avenues for exploring
other graph-related properties, e.g., connectivness and reachability in general.

\begin{small}
\section*{Acknowledgments}\vskip-2ex
This work was supported by national funds through Funda\c{c}\~{a}o para a Ci\^encia e a Tecnologia (FCT) with reference UID/CEC/50021/2013, and by the \emph{PrideMM Web Interface} grant from VeTSS\null.
\end{small}
\vskip-2ex

\bibliographystyle{short}
\bibliography{refs}
\end{document}